\documentclass[usenatbib]{mn2e}
\usepackage{psfig}
\usepackage{times}
\usepackage{amssymb}
\usepackage{graphicx}
\usepackage{graphics}

\def\apj{ApJ} \def\apjl{ApJL} \def\mnras{MNRAS} 
  \def\araa{ARAA} \def\aap{A\&AP}
\def\aj{AJ} \def\apjs{APJS}  \def\nat{Nature}

\def\gs{\mathrel{\raise0.35ex\hbox{$\scriptstyle >$}\kern-0.6em
\lower0.40ex\hbox{{$\scriptstyle \sim$}}}}
\def\ls{\mathrel{\raise0.35ex\hbox{$\scriptstyle <$}\kern-0.6em
\lower0.40ex\hbox{{$\scriptstyle \sim$}}}}

\def\kms{\,\hbox{km}\,\hbox{s}^{-1}}

\def\Wm2{\,\hbox{W}\,\hbox{m}^{-2}}
\def\gsim{\mathrel{\raise0.35ex\hbox{$\scriptstyle >$}\kern-0.6em\lower0.40ex\hbox{{$\scriptstyle \sim$}}}}
\def\lsim{\mathrel{\raise0.35ex\hbox{$\scriptstyle <$}\kern-0.6em\lower0.40ex\hbox{{$\scriptstyle \sim$}}}}
\def\ltsima{$\; \buildrel < \over \sim \;$}
\def\simlt{\lower.5ex\hbox{\ltsima}}
\def\gtsima{$\; \buildrel > \over \sim \;$}
\def\simgt{\lower.5ex\hbox{\gtsima}}

\def\etal{et~al.}

\begin{document}

\title[LARGE-SCALE OUTFLOW IN A $z\approx$~2 ULIRG]{Searching for Evidence of Energetic Feedback in Distant Galaxies: A Galaxy Wide Outflow in a $z\approx2$ Ultraluminous Infrared Galaxy}


\author[D.~M.~Alexander et al.]
{ \parbox[h]{\textwidth}{ 
D.\ M.\ Alexander,$^{\! 1\, *}$
A.\,M.\ Swinbank,$^{\! 2}$
Ian Smail,$^{\!2}$
R.\,McDermid,$^{\! 3}$
and N.\ P.\ H.\ Nesvadba$^{4}$
}
\vspace*{6pt} \\
$^1$Department of Physics, Durham University, South Road, Durham, DH1 3LE, UK \\
$^2$Institute for Computational Cosmology, Durham University, South Road, Durham, DH1 3LE, UK \\
$^3$Gemini Observatory, 670 N.\ A`ohoku Place, Hilo, HI 96720,
USA\\
$^4$Institut d'Astrophysique Spatiale, Universit\'e Paris Sud 11, Orsay, France\\
$^*$Email: d.m.alexander@durham.ac.uk \\
}
\maketitle
\begin{abstract}
Leading models of galaxy formation require large-scale energetic
outflows to regulate the growth of distant galaxies and their central
black holes. However, current observational support for this
hypothesis at high redshift is mostly limited to rare $z>2$ radio
galaxies. Here we present Gemini-North NIFS Intregral Field Unit (IFU)
observations of the [O~{\sc iii}]$\lambda$5007 emission from a
$z\approx$~2 ultraluminous infrared galaxy ($L_{\rm
IR}>10^{12}$~$L_{\odot}$) with an optically identified Active Galactic
Nucleus (AGN). The spatial extent ($\approx$~4--8~kpc) of the high
velocity and broad [O~{\sc iii}] emission are consistent with that
found in $z>2$ radio galaxies, indicating the presence of a
large-scale energetic outflow in a galaxy population potentially
orders of magnitude more common than distant radio galaxies. The low
radio luminosity of this system indicates that radio-bright jets are
unlikely to be responsible for driving the outflow. However, the
estimated energy input required to produce the large-scale outflow
signatures (of order $\approx10^{59}$~ergs over $\approx$~30~Myrs)
could be delivered by a wind radiatively driven by the AGN and/or
supernovae winds from intense star formation. The energy injection
required to drive the outflow is comparable to the estimated binding
energy of the galaxy spheroid, suggesting that it can have a
significant impact on the evolution of the galaxy. We argue that the
outflow observed in this system is likely to be comparatively typical
of the high-redshift ULIRG population and discuss the implications of
these observations for galaxy formation models.

\end{abstract}


\section{Introduction}

The most successful current models of galaxy formation invoke
large-scale energetic outflows to explain many of the properties of
local massive galaxies and the intragalactic medium (IGM; i.e.,\ red
optical colours of massive galaxies, steep optical galaxy luminosity
function, the black-hole--spheroid mass relationship, metal-enrichment
of the IGM; e.g.,\ Silk \& Rees 1998; Fabian 1999; King 2003; Benson
et~al. 2003; Granato et~al. 2004; Di Matteo et~al. 2005; Springel
et~al. 2005; Bower et~al. 2006; Croton et~al. 2006; Hopkins
et~al. 2006). A key attribute of these simulated outflows is the
injection of significant amounts of kinetic energy into the
interstellar medium (ISM), which can inhibit and terminate star
formation by either heating the ISM or ejecting the gas out of the
gravitational potential of the host galaxy. Large-scale outflows can
be powered by star formation and/or Active Galactic Nuclei (AGN)
activity (e.g.,\ Heckman et~al. 1990; Crenshaw et~al. 2003; Veilleux
et~al. 2005), although most galaxy formation models predict that only
AGN-driven outflows will be sufficiently energetic to have significant
impact on the formation and evolution of massive galaxies. These
large-scale outflows are expected to be particularly effective at
$z\approx$~2, when star formation in the most massive galaxies was in
significant decline (e.g.,\ Juneau et~al. 2005; Panter et~al. 2007;
P{\'e}rez-Gonz{\'a}lez et~al. 2008; Damen et~al. 2009).

Broadly speaking, AGN-driven outflows are kinematically energetic
``winds'' or ``jets'' that are initially launched close to the central
black hole. The two main catalysts that have been proposed to power an
AGN-driven outflow are (1) a radio jet or lobe, and (2) a radiatively
driven wind. A radio-jet/lobe driven outflow will be most prevalent in
radio-loud AGN but might only occur in a minority of the AGN
population.\footnote{The typical definition of a radio-loud AGN is
${\nu}{L_{\rm 5GHz}}$/${\nu}{L_{\rm 440nm}}\simgt$~10 (e.g.,\
Kellerman et al. 1989).} For example, $\approx$~4--24\% of the
optically bright (radiatively strong) AGN population are radio loud,
which appears to be a function of both redshift and AGN luminosity
(e.g.,\ Hooper et~al. 1995; Jiang et~al. 2007), although this may
indicate the ``duty cycle'' of radio AGN activity. The fraction of
optically faint (radiatively weak) radio-loud AGN fraction is a strong
function of stellar mass (e.g.,\ Best et~al. 2005; Smol{\v c}i{\'c}
et~al. 2009). By comparison, a radiatively driven wind would be most
effective in luminous AGNs with high mass-accretion rates, since the
wind is directly driven by radiation pressure from the accretion disk.

High signal-to-noise ratio X-ray and ultra-violet absorption-line
spectroscopy have shown that high-velocity outflows are present in
many AGNs, and may be a ubiqitious property of the AGN population at
both low and high redshift (e.g.,\ Crenshaw et~al. 1999: Chartas
et~al. 2002, 2007a,b; Laor \& Brandt 2002; Pounds et~al. 2003; Reeves
et~al. 2003; Porquet et~al. 2004; Ganguly \& Brotherton 2008; Gibson
et~al. 2009). For example, at least $\approx$~60\% of unobscured AGNs
show evidence for high-velocity outflows, and the maximum measured
outflow velocity is a strong function of the AGN luminosity (e.g.,\
Crenshaw et~al. 1999; Ganguly \& Brotherton 2008). The estimated
energy injection required to produce the X-ray absorption features
(which sometimes suggest outflow velocities exceeding
$v\approx$~0.1$c$) is typically $\approx$~0.1--1 of the AGN bolometric
luminosity (e.g.,\ Pounds et~al. 2003; Reeves et~al. 2003; Chartas
et~al. 2007a; Pounds \& Reeves 2009). However, X-ray variability
studies and consideration of the required energy input to produce the
high-velocity features, indicate that these outflow signatures must be
produced in the vicinity of the accretion disc on $<$~1~pc scales
(e.g.,\ Crenshaw et~al. 2003; King \& Pounds 2003). Therefore, while
it is clear that energetic outflows are present close to the accreting
black hole of AGNs, it is far from clear what impact these outflows
will have on the gas and star formation in the host galaxy on
$\approx$~1--10~kpc scales. To directly test the impact that outflows
have on the formation and evolution of galaxies, it is necessary to
identify large-scale energetic outflows, particularly in high-redshift
($z\simgt2$) massive galaxies where they are predicted to have been
most prevalent.

Optical spectroscopy of distant galaxies have revealed blueshifted
absorption/emission lines in many systems, a signature of outflowing
gas (e.g.,\ Pettini et~al. 2001; Shapley et~al. 2003; Tremonti
et~al. 2007). However, the most direct insight into the identification
and interpretation of large-scale outflows is provided by spatially
resolved spectroscopy (integral-field unit; IFU observations), which
provide direct information about both the velocity {\it and} extent of
any outflowing gas (e.g.,\ Swinbank et~al. 2005, 2006; Nesvadba
et~al. 2006). Using rest-frame optical IFU observations of distant
radio-loud AGNs (i.e.,\ high-redshift radio galaxies, hereafter
referred to as HzRGs), Nesvadba et~al. (2006, 2007a, 2008) showed that
large-scale energetic outflows are present in at least a fraction of
the high-redshift galaxy population. The key diagnostic that revealed
an AGN-driven outflow in these systems was the presence of
kinematically complex and extended [O~{\sc iii}]$\lambda$5007
emission;\footnote{Emission from [O~{\sc iii}] can be produced by a
number of processes, including photo-ionisation and shocks (Osterbrock
1989) but it cannot be produced in dense environments, such as the
broad-line region of AGN, without being collisionally de-excited
(e.g.,\ Osterbrock 1989; Robson 1996). The detection of broad
(FWHM$\simgt$~500--1000~km$^{-1}$) [O~{\sc iii}] emission therefore
indicates the presence of off-nuclear kinematic components.} see Holt
et~al. (2008) for similar constraints on $z\simlt0.6$ radio
galaxies. Nesvadba et~al. (2008) showed that the extent of the broad
[O~{\sc iii}] emission is similar to that of the radio emission,
providing evidence for a causal connection between the radio emission
and a large-scale outflow. The estimated kinetic energy required to
drive the outflows in these HzRGs is $\approx$~1--40\% of that
potentially provided by the radio jet, indicating that they could be
powered by mechanical energy from the radio jet. However, HzRGs are
rare systems (co-moving space densities at $z\approx$~2 of
$\Phi\approx10^{-8}$--$10^{-7}$~Mpc$^{-3}$; e.g.,\ Willott
et~al. 1998), possibly representing the most massive galaxies in the
most biased regions of the distant Universe (e.g.,\ Stevens
et~al. 2003; Seymour et~al. 2007). To fully assess the global impact
of large-scale energetic outflows on the formation and evolution of
distant galaxies, and to distinguish between the different mechanisms
proposed to drive outflows, it is therefore necessary to search for
large-scale energetic outflows in the more typical radio-quiet AGN
population.


In this paper we present Gemini-North NIFS IFU observations of the
[O~{\sc iii}] emission from a $z\approx$~2 quasar (SMM
J123716.01+620323.3, hereafter referred to as SMM~J1237+6203) that is
hosted in an Ultraluminous Infrared (IR) Galaxy (ULIRG; $L_{\rm
IR}>10^{12}$~$L_{\odot}$; e.g.,\ Sanders \& Mirabel 1996). ULIRGs
(both at low and high redshift) are rapidly evolving galaxies
undergoing intense dust-obscured star formation and AGN activity
(e.g.,\ Sanders \& Mirabel 1996; Genzel et~al. 1998; Page et~al. 2001;
Alexander et~al. 2005b; Stevens et~al. 2005; Schweitzer et~al. 2006;
Coppin et~al. 2008; Lutz et~al. 2008; Pope
et~al. 2008). Ultra-luminous infrared quasars are often believed to be
a key phase in the evolution of a ULIRG, where energetic outflows are
starting to shut down star formation in the host galaxy, before
revealing an unobscured quasar (e.g.,\ Sanders et~al. 1988; Canalizo
\& Stockton 2001; Granato et~al. 2004; Hopkins et~al. 2005a; Kawakatu
et~al. 2006). It is possible that every massive galaxy in the local
Universe underwent a ULIRG phase at some time over the past
$\approx$~13~Gyrs during which the galaxy spheroid and black hole were
predominantly grown (e.g.,\ Swinbank et~al. 2006; Alexander
et~al. 2008; Tacconi et~al. 2008). The [O~{\sc iii}] luminosity of
SMM~J1237+6203 is comparable to the HzRGs studied by Nesvadba
et~al. (2006, 2007a, 2008), suggesting similar intrinsic AGN
luminosities. However, importantly, SMM~J1237+6203 is $\approx$~3--4
orders of magnitude fainter at radio luminosities than the HzRGs. We
identify the signatures of a large-scale energetic outflow in this
ultraluminous infrared quasar, showing that these features are not
restricted to the distant radio-loud AGN population. We estimate the
energy input required to produce these features, assess what could be
responsible for driving this large-scale outflow, and discuss the
implications for galaxy formation models. We have adopted
$H_{0}=71\kms$, $\Omega_{M}=0.27$ and $\Omega_{\Lambda}=0.73$; in this
cosmology 1$''$ corresponds to 8.5\,kpc at $z=2.0$.


%
\section{Target, Observations, and Data Reduction}
\label{sec:obs_red}

\subsection{SMM~J1237+6203}

The Gemini-North NIFS observations of SMM~J1237+6203 are part of an
on-going NIFS program to identify large-scale energetic outflows in
$z\approx$~2 galaxies bright at submillimetre (submm)
wavelengths.\footnote{Programmes GN-2008A-Q-58 and GN-2009B-Q-21.} The
optical position of SMM~J1237+6203 is $\alpha_{2000}=$~12$^{\rm h}$
37$^{\rm m}$ 16\fs 0,
$\delta_{2000}=$~$+62^\circ$03$^{\prime}23\farcs2$ (Barger
et~al. 2003).

SMM~J1237+6203 has a submm flux of S$_{850}$=5.3$\pm$1.7\,mJy (Chapman
et~al. 2005), is identified with a moderately bright VLA radio
counterpart (S$_{1.4}$=164.0$\pm$11\,$\mu$Jy; Biggs \& Ivison 2006),
and has an optical spectroscopic redshift of $z=2.07$ (Barger et~al.
2003). The rest-frame 1.4~GHz luminosity
($\approx3\times10^{24}$~W~Hz$^{-1}$) suggests a rest-frame far-IR
luminosity of $\approx6\times10^{12}$~$L_{\odot}$, under the
assumption of the local radio-far-IR luminosity correlation (Alexander
et~al. 2005a; e.g.,\ Helou et~al.  1985; Condon 1992), identifying
SMM~J1237+6203 as a ULIRG. The detection of SMM~J1237+6203 at submm
wavelengths also suggests that it is bolometrically luminous
($\approx10^{13}$~$L_{\rm bol}$; e.g.,\ Chapman et~al. 2005). The
strong detection of SMM~J1237+6203 in the $\approx1\farcs5$ VLA map
(Biggs \& Ivison 2006) but the non-detection in a sensitive
(root-mean-square noise of 15.6~$\mu$Jy~beam$^{-1}$) $0\farcs5$ MERLIN
map suggests that the radio emission is not dominated by a bright AGN
core and is extended on scales of $>0\farcs5$ (R.~Beswick, priv
comm). This provides evidence that the radio luminosity (and therefore
the far-IR luminosity) may be dominated by intense star-formation
activity ($\approx$~1000~$M_{\odot}$~yr$^{-1}$; calculated following
Eqn.~4 in Kennicutt 1998 using the radio-derived far-IR luminosity).

The optical counterpart of SMM~J1237+6203 is bright ($R=20.2$) and
optical--near-IR spectroscopy reveals the rest-frame UV--optical broad
emission lines expected from a quasar (i.e.,\ Ly$\alpha$; N{\sc v};
C{\sc iv}; H$\delta$--H$\alpha$ with
FWHM~$\approx$~2100--2700~km~s$^{-1}$; Chapman et~al. 2005; Takata
et~al. 2006; Coppin et~al. 2008). Deep {\it Chandra} observations
reveal that SMM~J1237+6203 is luminous ($L_{\rm 2-10
keV}\approx10^{44}$\,erg\,s$^{-1}$) and possibly obscured ($N_{\rm
H}\approx3\times10^{22}$~cm$^{-2}$) in the X-ray band (Alexander
et~al. 2005a). The apparently contradictory evidence of both luminous
broad emission lines and X-ray absorption imply that either the X-ray
source and the broad-line gas are not co-spatial or that the X-ray
absorption is due to an ionised outflowing wind (e.g.,\ Crenshaw
et~al. 2003); the latter would provide direct evidence for an
AGN-driven outflow in the central regions, although higher
signal-to-noise ratio data is required for confirmation. On the basis
of the X-ray luminosity, the bolometric output of SMM~J1237+6203
appears to be dominated by star-formation activity (Alexander
et~al. 2005a), which is consistent with the bright submm flux and the
probable extended radio emission. However, we caution that we do not
have unambiguous evidence for star-formation activity in
SMM~J1237+6203; high signal-to-noise ratio spectroscopy would provide
more direct evidence (e.g.,\ the detection of CO molecular gas or
low-excitation star-formation emission lines/broad-band features in
mid-IR--mm spectroscopy). Low-resolution Subaru OHS spectroscopy shows
that SMM~J1237+6203 has luminous, broad, and extended [O~{\sc iii}]
emission (Takata et~al. 2006), providing evidence for the presence of
a large-scale outflow.

\begin{figure}
\includegraphics[width=6.5cm,angle=90]{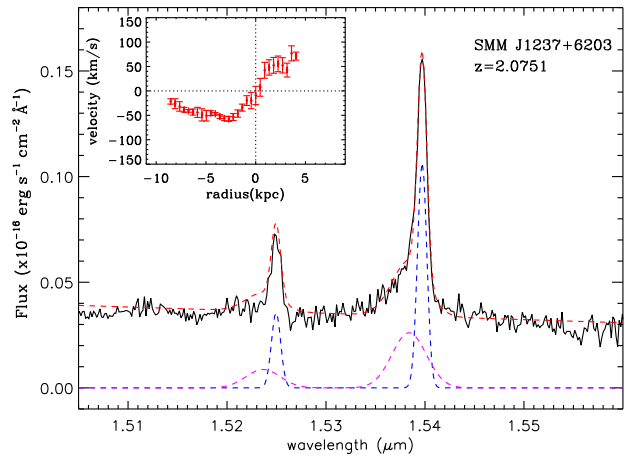}
\caption{Collapsed, one-dimensional integrated NIFS spectrum of
  SMM~J1237+6203. The spectrum is well fitted by a two component model
  with both a broad and narrow component (dotted curves show the best
  fitting emission-line profiles), suggesting the presence of an
  energetic outflow. The inset plot shows the narrow [O~{\sc iii}]
  velocity field extracted from the NIFS IFU datacube and plotted as a
  function of distance from the optically defined nucleus (region 4 in
  Fig.~2) along the major-axis.}
\label{fig:N24_onedspec}
\end{figure}

\subsection{Gemini-North NIFS Observations}

The NIFS IFU uses an image slicer to take a $3\farcs0\times3\farcs0$
field with a pixel scale of $0\farcs043$ and divides it into 29 slices
of width $0\farcs103$. The dispersed spectra from the slices are
reformatted on the detector to provide 2-dimensional
spectro-imaging. For our observations we used the $H$-band grism at a
spectral resolution of $\lambda/\Delta\lambda\approx$~5290
($\sigma\approx$~3\AA; $v\approx$~25\,km\,s$^{-1}$). They were taken
on 2008 May 22 and 2008 May 30 in excellent photometric conditions
with $\approx0\farcs3$ seeing. The observations were performed using
the standard ABBA configuration in which we chopped by 6$''$ to blank
sky to achieve good sky subtraction. Individual exposures were 600\,s
and the integrated exposure time was 15.6~ks (7.8~ks on-source and
7.8~ks on-sky).

%

We reduced the data with the Gemini {\sc iraf nifs} pipeline,
following the approach outlined in Swinbank et~al. (2009). The
data-reduction steps included sky subtraction, wavelength calibration,
and flat fielding. We removed residual OH sky emission using the
sky-subtraction routine described in Davies et~al. (2007). To
accurately align and and mosaic the individual datacubes, we created
white light (wavelength collapsed) images around the redshifted
[O~{\sc iii}] line from each observing block and centroided the galaxy
in the data cube. The data cubes were spatially aligned and co-added
to create the final mosaic using a median with a 3$\sigma$ clipping
threshold to remove remaining cosmetic defects and cosmic rays. Flux
calibration was carried out by observing standard stars at similar
airmass to the target galaxies, which were reduced in an identical
manner to the targets. Since the [O~{\sc iii}] emission lines lie in
regions of $>99$\% sky transparancy, no corrections were required for
telluric absorption. In all following sections, quoted line widths are
deconvolved for the instrumental resolution.

\begin{figure*}
\centerline{\psfig{figure=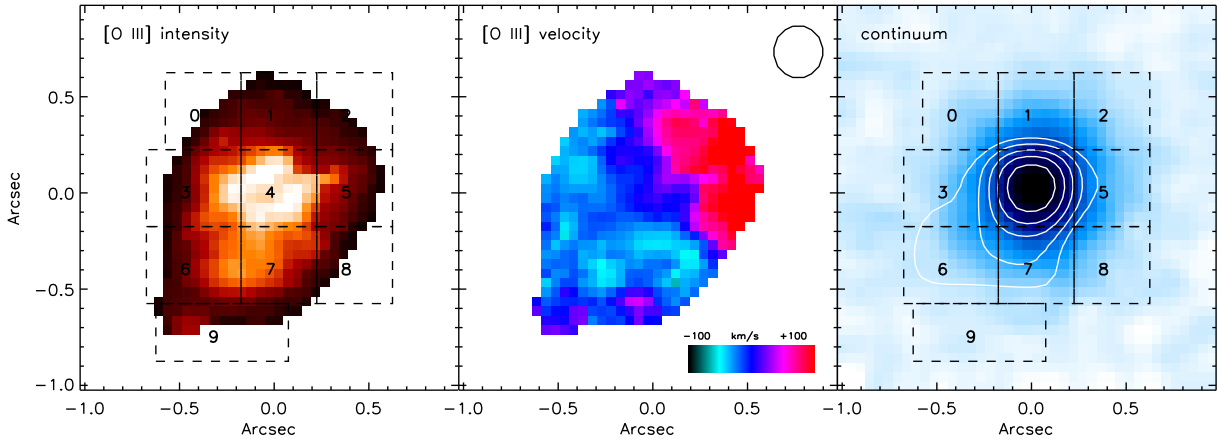,width=6.0in,angle=90}}
\caption{[O~{\sc iii}] intensity, narrow [O~{\sc iii}] velocity maps,
  and line-free continuum image (centred at 1.51~$\mu$m; rest-frame
  0.49~$\mu$m) of SMM~J1237+6203 from the NIFS data cube. North is up
  and East is left. The circle at the top right-hand corner of the
  middle panel denotes the seeing disk for the observations and the
  grid of dashed lines indicate the sub-regions where spectra are
  extracted; see Fig.~3. The [O~{\sc iii}]$\lambda$5007 emission is
  evidently complex, with clear velocity and intensity gradients. The
  contours in the right-hand panel represent an intensity-weighted map
  of the broad [O~{\sc iii}] emission components from Fig.~3, smoothed
  with a $\sigma$~=~2~pixel kernel (FWHM~=~4.75~pixels), which has an
  angular extent of $\approx0\farcs5$--$1\farcs0$
  ($\approx$~4--8~kpc).}
\label{fig:SMMJ1237_fitspec}
\end{figure*}

\begin{figure*}
\centerline{\psfig{figure=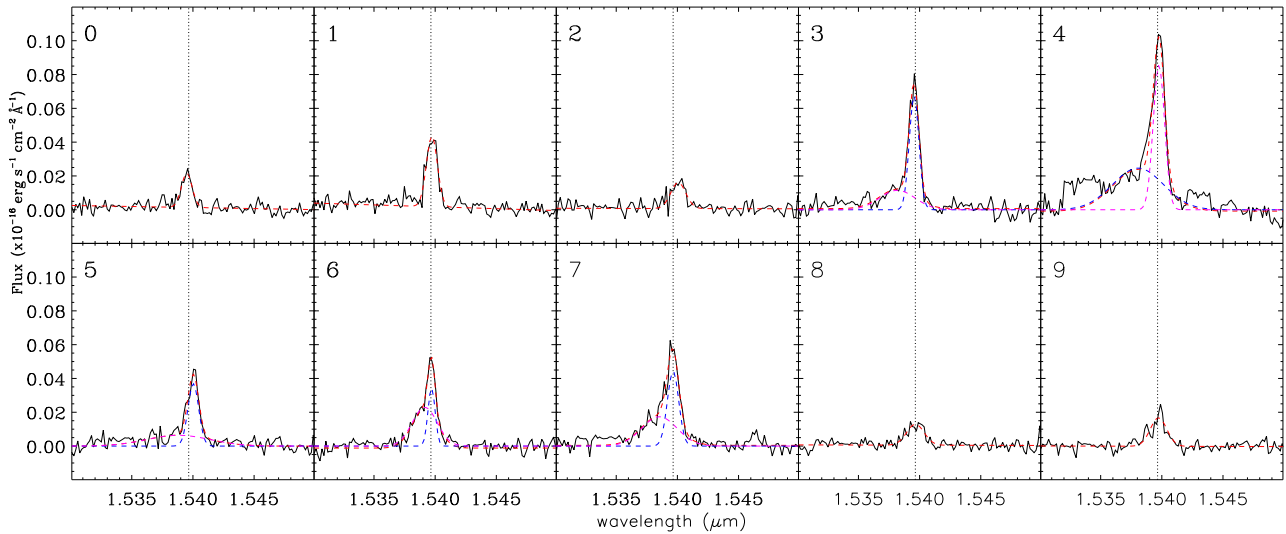,width=6.5in,angle=0}}
\vspace{0.3in}
\caption{Integrated spectra of SMM\,J1237+6203 for the ten sub-regions
  shown in Fig.~2. The red dashed curve denotes the best-fit
  emission-line parameters; in cases where a double Gaussian profile
  provide the best fit, both components are shown (dotted blue and
  magenta curves, offset in flux for clarity). The black dotted
  vertical line indicates the systemic redshift. The properties of the
  emission-line fits are given in Table~1.}
\label{fig:SMMJ1237_multispec}
\end{figure*}

\subsection{Emission-line Fitting}

We first constructed intensity and velocity maps of the [O~{\sc iii}]
emission from the IFU data cube. We used a $\chi^2$ minimisation
procedure to fit each spectrum within the datacube, taking into
account the greater noise at the positions of the sky lines. The
spectra were averaged over a $3\times3$ spatial pixel
($0\farcs15\times0\farcs15$), increasing to 5$\times$5 pixels and
ultimately to 7$\times$7 pixels if the signal was too low to give a
sufficiently high $\chi^2$ improvement over a fit without the line. In
regions where this averaging process still failed to give an adequate
$\chi^2$, no fit was made. To detect an emission line we required a
$\Delta\chi^2>25$ ($>5$~$\sigma$) improvement in a
continuum+emission-line fit over a simple continuum fit. When this
criterion is met we fit the [O~{\sc iii}] emission line with a single
Gaussian profile, allowing the normalisation, width and central
wavelength to vary.

Since both broad and narrow [O~{\sc iii}] emission-line components are
seen in the collapsed one-dimensional spectrum (see Fig.~1), we also
produced spatially binned spectra in 10 discrete regions to search for
the presence of broad [O~{\sc iii}] emission. For these integrated
spectra we fit both a single and double Gaussian profile emission line
but only accepted the fit if the double Gaussian profile resulted in a
significantly better fit than a single Gaussian profile
($\Delta\chi^2>16$ improvement; $>4$~$\sigma$). We present the
emission-line fitting results in Table~1.

\section{Results}
\label{sec:SMMJ1237}

\begin{table*}
{\scriptsize
\begin{center}
{\centerline{\sc SMM~J1237+6203 [O~{\sc iii}]$\lambda$5007 Emission-Line Constraints}}
\smallskip
\begin{tabular}{lccccccccc}
\hline
\noalign{\smallskip}
Region   & N$_{\rm comp}$ & $\Delta \chi^2$ & $z$ (narrow) & Flux (narrow)  & FWHM (narrow)  & $z$ (broad)   & Flux (broad)    & FWHM (broad)  & $\Delta v$ (broad)   \\
         &            &                 &             & ($10^{-16}$~erg~cm$^{-2}$~s$^{-1}$) & (km\,s$^{-1}$) & & ($10^{-16}$~erg~cm$^{-2}$~s$^{-1}$) & (km\,s$^{-1}$) & (km\,s$^{-1}$) \\
\hline
0        & 1          &   4             & 2.0747[2]   & 0.22[3]   & 160$\pm$35   & -          & -        & -            & -            \\
1        & 1          &   6             & 2.0751[2]   & 0.46[2]   & 165$\pm$35   & -          & -        & -            & -            \\
2        & 1          &   3             & 2.0746[5]   & 0.23[3]   & 244$\pm$47   & -          & -        &              & -            \\
3        & 2          &  45             & 2.0748[2]   & 0.66[5]   & 181$\pm$40   & 2.0721[15] & 0.42[6]  & 630$\pm$181   & 293$\pm$145  \\
4        & 2          & 150             & 2.0752[2]   & 0.95[10]  & 212$\pm$47   & 2.0719[16] & 1.28[18] & 935$\pm$212   & 312$\pm$150  \\
5        & 2          &  17             & 2.0760[3]   & 0.41[5]   & 132$\pm$49   & 2.0751[7]  & 0.41[6]  & 364$\pm$56   & 0$\pm$67    \\
6        & 2          &  45             & 2.0751[5]   & 0.25[7]   & 132$\pm$33   & 2.0740[12] & 0.58[5] & 468$\pm$85   & 107$\pm$116  \\
7        & 2          &  50             & 2.0750[4]   & 0.51[7]   & 209$\pm$42   & 2.0728[10] & 0.62[8] & 627$\pm$129   & 224$\pm$96   \\
8        & 1          &   2             & 2.0750[4]   & 0.22[3]   & 296$\pm$71   & -          & -        & -            & -            \\
9        & 1          &  3              & 2.0752[5]   & 0.28[2]   & 275$\pm$66   & -          & -        & -            & -            \\
\hline\hline
\label{table:SMMJ1237_multispec}
\end{tabular}
\caption{The best-fit parameters to the [O~{\sc iii}]$\lambda$5007
  emission line in SMM\,J1237+6203. Values in [] denote the error in
  the last decimal place. The $\Delta\chi^2$ statistic corresponds to
  the improvement in $\chi^2$ from fitting a double Gaussian
  emission-line profile over that of a single Gaussian emission-line
  profile. If the addition of the second Gaussian component gives a
  $\Delta\chi^2>16$ ($>4$~$\sigma$) improvement over the single
  Gaussian component fit, then we view the detection as significant
  and give the parameters of both (broad and narrow) emission
  lines. The regions (0--9) in column 1 denote those shown in Figs.~2
  \& 3. The $\Delta v$ velocity value corresponds to the offset
  between the systemic redshift and the wavelength of the broad
  [O~{\sc iii}] component in each integrated spectrum.}
\end{center}
}
\end{table*}

The collapsed one-dimensional integrated NIFS spectrum clearly shows
strong [O~{\sc iii}]$\lambda\lambda$4959,5007 emission; see
Fig.~1. The [O~{\sc iii}]$\lambda$5007 emission line has a blue
asymmetric profile, which is fitted with two underlying Gaussian
profiles (the line morphology is also displayed in the weaker [O~{\sc
iii}]$\lambda$4959 emission line). The narrow component of the [O~{\sc
iii}]$\lambda$5007 emission is centered at a redshift of
$z_N$~=~2.0751$\pm$0.0001 (hereafter defined as the systemic redshift)
with a width of FWHM~=~212$\pm$28\,km\,s$^{-1}$. The broad [O~{\sc
iii}] emission-line component is centered at $z_B$~=~2.0725$\pm$0.0003
(\hbox{${\Delta}v$~=~--254$\pm$60\,km\,s$^{-1}$} from the narrow
component) and has a width of FWHM~=~823$\pm$118~km~s$^{-1}$. Forty
five percent of the [O~{\sc iii}]$\lambda$5007 emission is from the
broad component, with an observed luminosity of $L_{\rm
[OIII]}$~=~(1.1$\pm$0.2)~$\times10^{43}$~erg~s$^{-1}$; the observed
narrow [O~{\sc iii}] luminosity is $L_{\rm
[OIII]}$~=~(1.4$\pm$0.2)~$\times10^{43}$~erg~s$^{-1}$. The broad
[O~{\sc iii}] luminosity of SMM~J1237+6203 is comparable to the HzRGs
studied by Nesvadba et~al. (2006, 2007a, 2008;
$\approx$~[1--8]~$\times10^{43}$~erg~s$^{-1}$).

In Fig.~2 we show the [O~{\sc iii}] intensity, narrow [O~{\sc iii}]
velocity (measured with respect to the systemic redshift), and
line-free optical continuum of SMM~J1237+6203, measured from the NIFS
data cube. The [O~{\sc iii}] emission is strongly peaked and
co-spatial with the optical continuum peak, as would be expected for
an optically luminous AGN. However, there is also a bright spatially
extended narrow velocity component ($\approx1\farcs7$;
$\approx$~14~kpc). The velocity map shows a strong gradient with a
peak-to-peak shift of 150$\pm$35\,km\,s$^{-1}$, with the highest and
lowest velocity gas to the South-East and North-West of the Nucleus,
respectively; see inset plot in Fig.~1. If it is assumed that the
velocity gradient of the [O~{\sc iii}] emission is due to host-galaxy
rotation then the dynamical mass estimated from the velocity curve is
$\approx10^{10}$\,M$_{\odot}$ for a canonical inclination of
$i=30^{\circ}$, but could be as high as $\approx10^{11}$\,M$_{\odot}$
for an inclination angle of $i=10^{\circ}$ (which is not potentially
unreasonable given that this object is a quasar and therefore likely
to be seen close to face on). However, due to potential complexities
from dust extinction and broad [O~{\sc iii}] components, it is not
clear that the [O~{\sc iii}] velocity field is dominated by galaxy
rotation; see Lehnert et~al. (2009) for other complications. By
comparison, the spheroid mass estimated from the velocity dispersion
of the narrow [O~{\sc iii}] emission ($\approx$~200~km~s$^{-1}$; see
also Table~1) is of order $\approx10^{11}$\,M$_{\odot}$ (calculated
following Eqn.~1 of Erb et~al. 2006 assuming $C=5$), which is
consistent with that expected from the estimated black-hole mass
($M_{\rm BH}\approx2\times10^{8}$~$M_{\odot}$; Alexander et~al. 2008)
given the local black-hole--spheroid mass relationship (e.g.,\
Tremaine et~al. 2002; Marconi \& Hunt 2003).

In Fig.~3 we show the [O~{\sc iii}] emission-line profile in the
integrated spectra from the ten discrete regions (labelled 0-9 in
Fig.~2). The [O~{\sc iii}] emission is evidently kinematically complex
in many of the regions across the galaxy. A broad
(FWHM~$\approx$~400--900~km~s$^{-1}$) [O~{\sc iii}] component is
identified in five ($\approx$~50\%) of the regions. The radial extent
of the broad [O~{\sc iii}] emission with respect to the narrow [O~{\sc
iii}] emission is shown in Fig.~2 and Fig.~4. Four of the broad
components (the exception is region 5, which has the lowest
$\Delta\chi^2$) have a velocity offset
($\Delta{v}\simgt0$~km~s$^{-1}$) with respect to the systemic redshift
of SMM~J1237+6203 and, interestingly, these regions are spatially
coherent (over $\approx$~4--8~kpc) and are located to the South East
of the nucleus; see Fig.~2. The presence of extended, high-velocity
broad [O~{\sc iii}] emission indicates a large-scale outflow, due
either to AGN and/or starburst activity. The lack of broad [O~{\sc
iii}] components to the North-West region of the nucleus could be due
to extinction by the host galaxy, which may be expected if the outflow
was driven by the AGN activity (i.e.,\ the far side of the outflow
would be obscured by the host galaxy). However, higher spatial
resolution and signal-to-noise ratio data is required to provide more
definitive spatial constraints.

\section{Discussion}

The Gemini-North NIFS observations of SMM~J1237+6203 show
high-velocity (up-to $\approx$~300~km~s$^{-1}$), broad (up-to
$\approx$~900~km~s$^{-1}$), large-scale ($\approx$~4--8~kpc) [O~{\sc
iii}] emission. The [O~{\sc iii}] properties of SMM~J1237+6203 are
consistent with those found for the HzRGs investigated by Nesvadba
et~al. (2006, 2007a, 2008) and indicate the presence of an energetic
large-scale outflow in SMM~J1237+6203. However, since SMM~J1237+6203
is $\approx$~3--4 orders of magnitude fainter at radio wavelengths
than HzRGs, it is unlikely that the catalyst for this outflow is a
radio jet.

SMM~J1237+6203 is the first high-redshift ULIRG with spatially
extended broad [O~{\sc iii}] emission to be mapped with IFU
observations. Previous IFU studies of high-redshift ULIRGs have
typically focused on the Ly~$\alpha$ or H~$\alpha$ emission line, and
the majority of the objects observed have not shown evidence for AGN
activity at optical wavelengths (e.g.,\ Bower et~al. 2004; Wilman
et~al. 2005; Swinbank et~al. 2005, 2006; Nesvadba
et~al. 2007b). However, three pieces of indirect evidence suggest that
SMM~J1237+6203 could be relatively typical of the high-redshift ULIRG
population: (1) broad [O~{\sc iii}] emission-line components have been
identified with rest-frame optical spectroscopy in several
high-redshift ULIRGs to date (e.g.,\ Smail et~al. 2003; Takata
et~al. 2006; Coppin et~al. 2008), (2) $\approx$~50\% of nearby ULIRGs
hosting optical AGN activity have [O~{\sc iii}] components with
FWHM~$>$~800~km~s$^{-1}$, comparable to that found for SMM~J1237+6203
(e.g.,\ Veilleux et~al. 1999; Zheng et~al. 2002; see also Spoon \&
Holt 2009 for similar mid-IR spectral constraints), and (3)
high-quality IFU data have been published for a number of nearby
ULIRGs showing that they host broad and extended [O~{\sc iii}],
providing evidence for large-scale energetic outflows in at least some
ULIRGs in the local Universe (e.g.,\ Colina et~al. 1999; Wilman
et~al. 1999; L{\'{\i}}pari et~al. 2009a,b).

In this section we explore the properties of the large-scale outflow
in SMM~J1237+6203 in more detail by, first, estimating the energy
input required to produce the observed outflow features (see \S4.1)
and, second, exploring what proceses could deliver sufficient energy
to drive the outflow (see \S4.2). Lastly, we consider the potential
fate of the outflowing gas and briefly discuss the implications of our
observations for galaxy formation models (see \S4.3).

\subsection{The Properties of the Large-Scale Outflow}

In Fig.~5 we plot the velocity offset and FWHM of the broad [O~{\sc
iii}] emission for SMM~J1237+6203 and compare it to HzRGs and samples
of $z<0.5$ quasars. We have not corrected the widths and velocities of
the [O~{\sc iii}] emission-line gas for unknown projection effects and
dust extinction, and therefore they should be considered lower
limits. There appears to be a loose correlation between the width of
the [O~{\sc iii}] emission and the [O~{\sc iii}] velocity offset,
which would be expected for outflowing gas that becomes turbulent as
the kinetic energy is dissipated through shocks into the ISM (e.g.,\
Lehnert \& Heckman 1996); however, we caution that this is unlikely to
be the only possible interpretation. This figure shows that
SMM~J1237+6203 has [O~{\sc iii}] properties broadly similar to those
of the HzRGs, with the exception of the most extreme source
(PKS~1138-262) which has very broad [O~{\sc iii}]; however, we have
only plotted the integrated [O~{\sc iii}] properties for the HzRGs
here and note that the widths and velocity offsets of spatially
distinct regions are often higher. This figure also shows that the
integrated [O~{\sc iii}] properties of SMM~J1237+6203 are (1) not as
extreme as some nearby ultraluminous infrared quasars, and (2) are
more extreme than the majority of nearby quasars; indeed, only an
extreme subset of the rare Narrow Line Seyfert 1 population (NLS1) are
found to host [O~{\sc iii}] properties similar to SMM~J1237+6203
(e.g.,\ on the basis of Komossa et~al. 2008, only $\approx$~5--10\% of
NLS1s have high-velocity [${\Delta}v\simgt200$~km~s$^{-1}$] broad
[O~{\sc iii}]). This further suggests that the origin of the outflow
features in SMM~J1237+6203 are coupled to the production of the
infrared emission, which we explore in more detail in \S4.2. However,
we stress that the [O~{\sc iii}] properties for the $z<0.5$ quasar
samples are from long-slit spectroscopy and provide limited
constraints on the spatial extent of these potential outflows.

The extent of the broad [O{\sc iii}] emission-line gas from
SMM~J1237+6203 is comparable to that found for the HzRGs studied by
Nesvadba et~al. (2006, 2007a, 2008), which typically range from
$\approx$~4--20~kpc, suggesting comparable overall energetics given
the similar widths and velocities of the broad [O~{\sc iii}]
emission. Assuming that the broad [O~{\sc iii}] properties are due to
an energy-conserving bubble expanding into a uniform medium (see
Eqn.~3 in Nesvadba et~al. 2006), we can place first-order constraints
on the kinetic energy required to accelerate the broad [O~{\sc iii}]
emission-line gas to these velocities over large distances. For
example, the estimated kinetic energy required to accelerate the broad
[O~{\sc iii}] emission to ${\Delta}v\approx$~300~km~s$^{-1}$ over
$\approx$~4--8~kpc (the estimated size of the broad [O~{\sc iii}]
emission; see Fig.~2) is of order
$\approx$~(0.6--3)~$\times10^{44}$~erg~s$^{-1}$ for SMM~J1237+6203,
which should be compared to
$\approx$~(0.3--100)~$\times10^{44}$~erg~s$^{-1}$ for the HzRGs; in
this calculation we assumed the same ambient density as that estimated
for the HzRGs by Nesvadba et~al. (2008) since we do not have direct
constraints for SMM~J1237+6203.

\begin{figure}
\includegraphics[angle=0,width=3.5in]{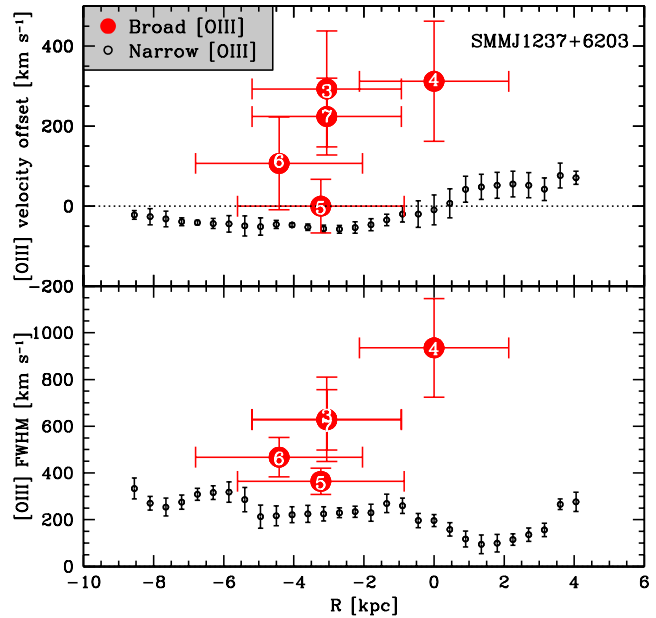}
\caption{Velocity and FWHM components of the broad and narrow [O~{\sc
iii}] emission of SMM~J1237+6203 plotted as a function of radius from
the optically defined nucleus (region 4 in Fig.~2): the regions where
broad [O~{\sc iii}] emission is detected are labelled. The radius and
bin width of the broad [O~{\sc iii}] components are determined from
Fig.~2 while the plotted narrow [O~{\sc iii}] component is the narrow
[O~{\sc iii}] velocity field shown in Fig.~1.}
\label{fig:oiiiradio}
\end{figure}

\begin{figure}
\includegraphics[angle=0,width=3.5in]{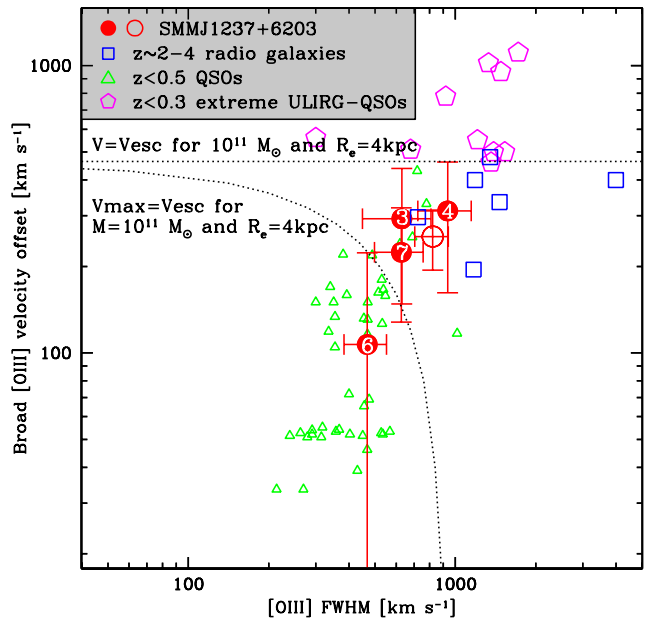}
\caption{Broad [O~{\sc iii}] emission-line velocity offset versus FWHM
found for SMM~J1237+6203, the HzRGs from Nesvadba et~al. (2006, 2007a,
2008), $z<0.3$ ultra-luminous infrared quasars (Zheng et~al. 2002),
and samples of $z<0.5$ quasars (Boronson et~al. 2005; Komossa
et~al. 2008). The data for the nearby samples is from long-slit
spectroscopy and provide limited constraints on the extent of the
[O~{\sc iii}] emission. The [O~{\sc iii}] emission from the HzRGs is
extended on $\simgt1$~kpc scales but here we only plot the collapsed
one-dimensional IFU spectra; however, for SMM~J1237+6203 we show the
data from the collapsed one-dimensional IFU spectrum (open circle) and
the four individual regions that have broad [O~{\sc iii}] emission
with ${\Delta}v\simgt0$~km~s$^{-1}$ (filled circles): the individual
regions are labelled. The maximum velocities ($v_{\rm
max}$~=~$v$~+~1/2~FWHM; i.e.,\ Martin 2005; Rupke et~al. 2005a,b) of
the broad emission-line gas for SMM~J1237+6203 are potentially high
enough to exceed the escape velocity ($v_{\rm esc}$) of a massive
galaxy (dotted curve), although since the majority of the gas has
$v<v_{\rm esc}$ it seems likely that only a minority of the gas will
be expelled from the host galaxy.}
\label{fig:oiiiradio}
\end{figure}

Over a canonical 30~Myr quasar lifetime (e.g.,\ Martini \& Weinberg
2001; Hopkins et~al. 2005b), the total injection of energy into the
outflow would be of order $\approx$~(0.3--3)~$\times10^{59}$~ergs,
which is comparable to the estimated binding energy of the galaxy
spheroid in SMM~J1237+6203; for example, based on an estimated
spheroid mass of $\approx10^{11}$~$M_{\odot}$ and velocity dispersion
of $\sigma\approx200$~km~s$^{-1}$ (see \S3), the estimated binding
energy is $\approx10^{59}$~ergs for $R_{\rm e}=4$~kpc (e.g.,\ Binney
\& Tremaine 1987). This analysis is based on a simple model and should
only be considered illustrative with uncertainties at the level of an
order of magnitude but, given the limited constraints current
available for high-redshift systems, a more complex model is not yet
warranted. However, it does indicate that the large-scale outflow in
SMM~J1237+6203 may be energetic enough to unbind at least a fraction
of the ISM from the host galaxy.

\subsection{The Potential Power Sources of the Large-Scale Outflow}

On the basis of a comprehensive analysis of the broad [O~{\sc iii}]
properties of HzRGs, Nesvadba et~al. (2006, 2007a, 2008) argued that
the outflows observed in HzRGs are most likely powered by radio jets,
which can potentially inject of order $\approx10^{46}$~erg~s$^{-1}$ of
mechanical energy into the ISM (although this calculation is dependent
on the assumed radio luminosity to kinetic jet-power ratio). The
kinematics, extent, and luminosities of the broad [O~{\sc iii}]
emission from SMM~J1237+6203 are comparable to those of the
HzRGs. However, since the 1.4~GHz radio luminosity of SMM~J1237+6203
is $\approx$~3--4 orders of magnitude lower than that found for the
HzRGs, this strongly suggests that mechanical energy from a radio jet
is not the catalyst for the broad [O~{\sc iii}] emission seen in
SMM~J1237+6203. For example, taking the same assumptions as those used
in Nesvadba et~al. (2006), we predict the mechanical energy from a
radio jet in SMM~J1237+6203 to be of order
$\approx10^{42}$~erg~s$^{-1}$, which is two orders of magnitude below
the required energy input to produce the observed broad [O~{\sc iii}]
features, even assuming a 100\% energy--ISM coupling efficiency. This
estimate of the radio-jet mechanical energy is also likely to be an
upper limit since it is possible that a large, possibly dominant,
fraction of the radio emission from SMM~J1237+6203 is due to star
formation rather than AGN activity (see \S2.1).  However, could the
high-velocity, broad, large-scale [O~{\sc iii}] emission from
SMM~J1237+6203 be powered by an outflow from a radiatively driven AGN
wind or, given the indirect evidence for star-formation activity (see
\S2.1), supernova winds associated with an intense starburst?

Estimating the potential impact of an radiatively driven AGN wind on
the ISM is difficult due to the huge difference in size scale between
the likely $<1$~pc ejection radius of the wind and the large-scale
($>1$~kpc) galactic enviroment. However, given the constraints from
high signal-to-noise ratio X-ray spectroscopy for a number of AGNs to
date (see \S1), it seems likely that a significant fraction of the
bolometric luminosity from the AGN will drive an outflow close to the
accretion disk, which at least provides a first-order constraint on
the {\it initial} energy injection. Given the low signal-to-noise
ratio of the X-ray spectra for SMM~J1237+6203, we cannot directly
identify the X-ray signatures of a high-velocity outflow in these
data. However, we note that on the basis of high signal-to-noise ratio
X-ray spectral analyses of the gravitationally lensed \hbox{$z=3.9$}
ultraluminous infrared quasar APM~08279+5255 (potentially a
higher-redshift analog of SMM~J1237+6203), we may expect $\approx$~0.1
of the AGN bolometric luminosity of SMM~J1237+6203 to be injected into
an outflow (e.g.,\ Chartas et~al. 2007a; see also Chartas et~al. 2007b
for similar X-ray spectral constraints on another distant
ultraluminous infrared quasar); this energy injection fraction is also
similar to that typically assumed in theoretical simulations (e.g.,\
Wyithe \& Loeb 2003; Di Matteo et~al. 2005). The total bolometric
luminosity of SMM~J1237+6203 is $\approx3\times10^{46}$~erg~s$^{-1}$,
although a significant fraction of this emission may be due to
star-formation activity; see \S2.1. A more conservative constraint on
the AGN bolometric luminosity is $\approx3\times10^{45}$~erg~s$^{-1}$,
based on the rest-frame 2--10~keV luminosity and the average spectral
energy distribution for quasars from Elvis et~al. (1994). The {\it
initial} energy input into the accretion-disk wind is therefore likely
to be $\approx$~(0.3--3)~$\times10^{45}$~erg~s$^{-1}$, which would
correspond to a total energy injection over a 30~Myr quasar lifetime
of $\approx$~(0.3--3)~$\times10^{60}$~ergs. Therefore, so long as
$\approx$~10--100\% of the energy from the wind can couple to the
large-scale ISM gas, then the quasar can drive the outflow to
$\approx$~8~kpc (the maximum extent; see Fig.~2); the wind--ISM gas
coupling efficiency drops to $\approx$~2--20\% to drive the outflow to
$\approx$~4~kpc (the minimum extent; see Fig.~2). These wind--gas
coupling efficiencies are plausible given the likely large solid angle
of radiatively driven AGN winds (e.g.,\ Proga et~al. 2000; King \&
Pounds 2003; Ohsuga et~al. 2009).


The dominant source of energy injection from intense star-formation
activity is supernovae winds. Dalla Vecchia \& Schaye (2008) estimate
the energy input from supernovae to be
$\approx$~$10^{49}$~erg~$M_{\odot}^{-1}$, under the assumption of a
Salpeter initial mass function. Assuming a constant star-formation
rate of $\approx$~1000~$M_{\odot}$~yr$^{-1}$ (see \S2.1), the energy
injection into the host galaxy from supernovae winds would be
$\approx3\times10^{44}$~erg~s$^{-1}$. This suggests that the
large-scale energetic outflow observed in SMM~J1237+6203 could be
powered by intense star-formation activity so long as
$\approx$~20--100\% of the injected supernovae energy can be coupled
to the ISM (the range in wind-gas coupling fractions correspond to the
range in [O~{\sc iii}] extent of 4--8~kpc; see Fig.~2). Given that the
{\it energy injection} from supernovae will be more widespread than
that from the AGN (i.e.,\ potentially over kpc scales rather than pc
scales), such a large coupling fraction may not be unreasonable
(e.g.,\ Weaver et~al. 1977). Furthermore, our calculation of the
required energy to produce the broad [O~{\sc iii}] features was based
on the assumption that the outflow needed to be driven over
$\approx$~4--8~kpc scales, when the distance could be substantially
lower if the star-formation activity is widespread. For example, the
energy injection required to produce the broad [O~{\sc iii}] features
over $\approx$~1--2~kpc scales is about an order of magnitude lower
than that required to produce the broad [O~{\sc iii}] features over
$\approx$~4--8~kpc scales, which would give similar coupling fractions
as those found in nearby starburst-dominated ULIRGs (e.g.,\ Martin
2006). This suggests that even if the star formation in SMM~J1237+6203
is lower than that assumed here then it is still possible to produce
the broad [O~{\sc iii}] features so long as the supernova winds can
efficiently couple to the ISM.

Both an AGN-powered and star-formation-powered wind are plausible
candidates for driving the large-scale outflow in SMM~J1237+6203, and
on the basis of our current constraints we cannot directly tell which
process dominates. As mentioned above, broad [O~{\sc iii}] components
with FWHM~$>$~800~km~s$^{-1}$ are detected in $\approx$~50\% of nearby
ULIRGs hosting optical AGN activity (e.g.,\ Veilleux et~al. 1999;
Zheng et~al. 2002) while, by comparison, no nearby ULIRGs optically
classified as HII galaxies or LINERs are found to host broad [O~{\sc
iii}] emission. This would seem to suggest that the production of
broad [O~{\sc iii}] is related to AGN activity. However, since [O~{\sc
iii}] emission is easily photoionised by AGN activity, the detection
of broad [O~{\sc iii}] components in optically identified AGNs could
be a sensitivity/selection effect (i.e.,\ the AGN may just be more
effective at illuminating the outflowing gas that is potentially
present in all ULIRGs). Indeed, optical spectroscopy of the Na~ID
absorption line has shown that $\simgt$~50\% of ULIRGs host outflowing
neutral gas, irregardless of the optical spectral type (i.e.,\ HII,
LINER, Seyfert classification), indicating that star-formation
activity may often be responsible for gaseous outflows in ULIRGs
(e.g.,\ Martin 2005; Rupke et~al. 2005a,b). It is therefore also
interesting that typical quasars (i.e.,\ those that are not infrared
bright) rarely show such broad [O~{\sc iii}] features (e.g.,\ Boroson
et~al. 2005; Komossa et~al. 2008; see Fig.~5), which may reflect
either an absence of intense star-formation activity or different
evolutionary stages (i.e.,\ an ultraluminous infrared quasar may be an
earlier phase associated with energetic outflows; see \S1). However,
all of the evidence presented here is indirect and sensitive IFU
observations of larger samples of distant ULIRGs, AGNs, and
star-forming galaxies are required to provide more direct constraints.

\subsection{The Potential Fate of the Outflowing Gas}

A crucial feature of large-scale energetic outflows in galaxy
formation models is the potential to eject gas from the host
galaxy. Although we have shown that the energy input from both AGN and
star-formation activity is potentially large enough to drive the
large-scale outflow and unbind the ISM, we have not investigated
whether the gas can be ejected from the host galaxy. Murray
et~al. (2005) show that a significant fraction of the ISM can only be
expelled from the host galaxy if the energy injection exceeds a
limiting luminosity, which is a function of the gas-mass fraction and
the galaxy gravitational potential (this limiting luminosity is
analogous to the Eddington limit for the galaxy; see Eqn. 18 in Murray
et~al. 2005). We do not have an accurate constraint on the mass of the
cold-molecular ISM gas in SMM~J1237+6203; however, if SMM~J1237+6203
is similar to typical submillimetre-emitting galaxies then the
moleculear gas mass would be of order
$\approx10^{10}$--$10^{11}$~$M_{\odot}$ (e.g.,\ Greve et~al. 2005),
which gives a gas-mass fraction $\approx$~0.1--1.0. Assuming
$\sigma$~=~200~km~s$^{-1}$ and a gas-mass fraction of 0.1--1.0, the
limiting luminosity for SMM~J1237+6203 is
$\approx$~(0.8--7.8)~$\times10^{13}$~$L_{\odot}$, which is comparable
to or larger than the total bolometric luminosity. Therefore, on the
basis of this analysis, it seems unlikely that the majority of the ISM
in SMM~J1237+6203 will be expelled from the host galaxy. The
velocities of the broad [O~{\sc iii}] emission are also in broad
agreement with this analysis, showing that while $v_{\rm
max}$$>$$v_{\sc esc}$ in some of the regions, the typical velocities
of the gas are more modest ($v<v_{\sc esc}$); see Fig.~5. In this
respect, SMM~J1237+6203 is similar to nearby starburst galaxies, which
have maximum outflowing gas velocities within a tight range between
the circular velocity and escape velocity of the host galaxy (e.g.,\
Martin 2005).

We therefore find that SMM~J1237+6203 is driving a large-scale
energetic outflow that is disrupting the ISM and could potentially
shut down star formation in the host galaxy, in qualitative agreement
with predictions from models of the rapid growth phase of galaxies and
their central black holes (e.g.,\ Granato et~al. 2004; Di Matteo
et~al. 2005; Hopkins et~al. 2006). If SMM~J1237+6203 represents a
typical ULIRG phase that all massive galaxies have undergone at some
time over the past $\approx$~13~Gyrs, then this outflow activity is
potentially $\approx$~2--3 orders of magnitude more common than that
found in HzRGs (i.e.,\ the comoving space density of $z\approx$~2
submm-bright galaxies is $\Phi\approx10^{-5}$~Mpc$^{-3}$ to be
compared to $\Phi\approx10^{-8}$--$10^{-7}$~Mpc$^{-3}$ for
$z\approx$~2 HzRGs; Willott et~al. 1998; Chapman et~al. 2005; Swinbank
et~al. 2006). Indeed, we have identified high signal-to-noise broad
[O~{\sc iii}] components in all of the three high-redshift ULIRGs
observed in our on-going Gemini NIFS programs so far. It is therefore
likely that the dominant outflow mechanism in distant galaxies is that
found in SMM~J1237+6203 (either AGN accretion-disk or supernovae
winds) rather than radio-jet induced AGN activity. However, since the
outflow observed in SMM~J1237+6203 requires significant gas accretion
to drive either an AGN or supernova wind, once the quasar and
star-formation activity has terminated the outflowing gas will cool,
which will initiate further star formation. Therefore, unless the gas
entropy in the system has become so high that the gas-cooling
timescales are $>1$~Gyr (e.g.,\ I.~McCarthy et~al. in prep), it seems
likely that another process (e.g.,\ a radiatively weak radio-loud AGN
as is commonly seen in nearby massive galaxies; Best et~al. 2005,
2006) is required to keep the gas unbound and to prevent further star
formation from occuring on comparatively short timescales.

\section{Conclusions}

We have presented Gemini-North NIFS IFU observations of the [O~{\sc
iii}]$\lambda$5007 emission in a $z\approx$~2 ultraluminous infrared
galaxy hosting an optically identified AGN (SMM~J1237+6203). Our main
findings are the following:

\begin{itemize}

\smallskip
\item The spatial extent ($\approx$~4--8~kpc) of the high velocity
(up-to $v\approx$~300~km~s$^{-1}$) and broad (up-to
$\approx$~900~km~s$^{-1}$) [O~{\sc iii}] emission is consistent with
that found in $z>2$ radio galaxies, suggesting the presence of a
large-scale energetic outflow in a galaxy population potentially
orders of magnitude more common than $z>2$ radio galaxies. See \S3,
\S4.1, \& \S4.3.

\smallskip
\item The estimated energy required to produce these large-scale
outflow features ($\approx$~[0.6--3]~$\times10^{44}$~erg~s$^{-1}$)
could be provided by a wind radiatively driven by the quasar
($\approx$~[0.3--3]~$\times10^{45}$~erg~s$^{-1}$) and/or supernovae
winds from intense star formation
($\approx3\times10^{44}$~erg~s$^{-1}$), so long as the wind-gas
coupling efficiencies are comparatively high (of order
$\approx$~10--100\%). However, the low radio luminosity of this system
indicates that radio-bright jets are unlikely to be responsible for
driving the outflow, in contrast to $z>2$ radio galaxies. See \S4.1
and \S4.2.

\smallskip
\item The estimated energy injection required to drive the large-scale
outflow is comparable to the estimated binding energy of the galaxy
spheroid in SMM~J1237+6203, suggesting that the outflow can have a
significant impact on the evolution of the galaxy. However, on the
basis of the maximum energy input into the outflow and the measured
outflow velocities, it seems unlikely that the majority of the ISM
will be expelled from the host galaxy. This may mean that another
process (or longer lived AGN activity) is required to keep the gas
unbound and prevent future star formation from occuring. See \S4.3.

\end{itemize}

Finally, we conclude that SMM~J1237+6203 is the first high-redshift
ULIRG with spatially extended broad [O~{\sc III}] emission to be
mapped with IFU observations, revealing the signatures expected for a
large-scale energetic outflow. However, since $\approx$~50\% of nearby
ULIRGs hosting optical AGN activity also have broad [O~{\sc iii}]
emission, we expect SMM~J1237+6203 to be comparatively typical of the
high-redshift ULIRG population. High signal-to-noise ratio (and high
spatial resolution) IFU observations of further high-redshift ULIRGs
will provide more definitive constraints.

\section*{acknowledgments}

We gratefully acknowledge support from the Royal Society (DMA), the
Leverhulme Trust (DMA), the Royal Astronomical Society (AMS), and the
Science and Technology Facilities Council (STFC; IRS). We thank the
anonymous referee for a detailed reading of the paper. We also thank
Rick Davies for proving his sky-subtraction code, Bob Beswick and Rob
Ivison for providing interpretation of the high-resolution MERLIN
radio data, David Rupke for providing spectrocopic data on local
ULIRGs, and Richard Bower for interesting discussions. This paper was
based on observations obtained at the Gemini Observatory, which is
operated by the Association of Universities for Research in Astronomy,
Inc., under a cooperative agreement with the National Science
Foundation (NSF) on behalf of the Gemini partnership: the NSF (United
States), the STFC (United Kingdom), the National Research Council
(Canada), CONICYT (Chile), the Australian Research Council
(Australia), Minist{\'e}rio da Ci{\^e}ncia e Tecnologia (Brazil) and
SECYT (Argentina).


{}

\end{document}